(Comment on "Interplanetary coronal mass ejections that are undetected by solar coronagraphs" by T. A. Howard and G. M. Simnett)

**Abstract**

Howard and Simnett (hereinafter referred to as HS) employed a new technique for associating LASCO CMEs to SMEI ICMEs. In order to extrapolate the SMEI data back to the LASCO field of view they used non-linear trajectories, dependent on a speed and direction, what is more realistic than the linear extrapolation with only one parameter (a speed). However, there are two errors and one mistake in their procedure: (1) HS used two free parameters of the direction, whereas only one can be freely selected, because the second is provided by SMEI data. As a result, the directions determined by HS are incorrect. (2) HS overlooked that, since the trajectory depends on more than one parameter, there is a broad set of trajectories, for various speeds and directions, matching the event, and thus a broad range of the onset times. HS select only one trajectory for each SMEI event. Therefore the associations made by them are incomplete, and they should be reexamined. As long as it is not done any conclusion about CMEs "undetected by solar coronagraphs" are premature. (3) HS made some mistake in determination of the SMEI speeds. The speeds given in their Table 1 are about twice as high as those demanded to obtain the onset times given in the table. It explains why the SMEI speed distribution is "excessively shifted toward high speeds"; hence, there is no reason to search for a physical explanation.

(Comment on "Interplanetary coronal mass ejections that are undetected by solar coronagraphs" by T. A. Howard and G. M. Simnett)

**History of the comment**

15 October 2008 – The comment has been submitted to Journal of Geophysical Research, Space Physics.

17 December 2008 – I received an editor's decision that the comment cannot be accepted for publication in JGR.

An attached review contains nothing but generalities. The reviewer supports his/her recommendation on the comment rejection with two arguments. Firstly, he/she claims that "the multi-valued nature" of the HS's technique is discussed in the original paper, but he/she does not indicate where it is done. However, there is no such discussion in HS. On the contrary, HS consider only one value of the SMEI onset time (see the second column of their Table 1).

The second argument used by the reviewer is that "the Comment fails to show that the HS conclusions are in error". However, I do not say in my comment that the conclusions are erroneous, but that associations made by HS are incomplete, so, insufficient to substantiate their conclusion.

Thus, the first argument is based on untruth, the second is missed. On the other hand, the reviewer totally ignored important question raised in my comment, namely, that the directions and speeds determined by HS are incorrect. Such results are misleading and they shouldn't be left without any comment.

I raised the above questions in a letter to the editor, but he did not change his decision.

**Comment on "Interplanetary coronal mass ejections that are undetected by solar coronagraphs" by T. A. Howard and G. M. Simnett**

Marta Skirgiello
Space Research Center of Polish Academy of Sciences, Warsaw, Poland

1. Introduction

Recently, Howard and Simnett [2008] (hereinafter referred to as HS) reported that some interplanetary coronal mass ejections (ICMEs) observed with the Solar Mass Ejection Imager (SMEI) had not counterparts in the LASCO observations. In order to associate a LASCO CME to the SMEI ICME they determine the onset time of the event using a linear fit extrapolation along with a new one called the Cube-Fit procedure. They demonstrated that elongation of SMEI ICMEs never linearly increases with time, because it is dependent not only on a speed, but also on a direction of the event. The authors developed a data cube containing elongation-time plots for various speeds and directions. For each SMEI event, they select the best-fit plot from the data cube; then they associate a CME from the LASCO data. Such method largely progresses the technique used in the previous papers. However, there are two sources of errors overlooked by the authors.

Firstly, HS use three independent parameters in the model, while two of them (latitude and longitude) are related by the position angle (PA) provided by the SMEI observation. One of the consequences is that the determined latitudes and longitudes are inconsistent with the measured position angle. This question is discussed in the next section.

Secondly, since the model has two free parameters (a speed and direction), there may be many elongation-time plots, with very different speeds and directions, matching a given event equally well. HS's procedure selects only one plot, thus their associations are incomplete, so, insufficient to qualify some ICMEs as 'weak associated' or 'missing LASCO data'. In consequence, main conclusion of their paper that there were SMEI ICMEs undetected by solar coronagraphs should be questioned. It is discussed in section 3 of the comment.

**2. Too many parameters of the model**

HS determine ICME direction using two parameters: the latitude ($\Lambda$) and longitude ($\Phi$). There is another pair of parameters which can be used instead of ($\Lambda,\Phi$), namely, the position angle (PA) and the center disk distance ($\Psi$). The parameters of the direction are shown in Fig. 1. Relations between them can be easily derived from the right spherical triangle (marked by thick line in the figure) where $\Lambda$ and $\Phi$ are legs, $\Psi$ is a hypotenuse, the angle $\lambda$, subtended the side $\Lambda$, is so called apparent latitude. The angle $\lambda$ is unambiguously determined by the PA.

From the spherical triangle in Fig. 1 you can obtain the relations used in this section:

$$\tan \Lambda = \sin \Phi \cdot \tan \lambda = \sin \Phi \cdot |\cot PA| \qquad (1)$$
$$\sin \Lambda = \sin \Psi \cdot \sin \lambda = \sin \Psi \cdot |\cos PA| \qquad (2)$$
$$\tan \Phi = \tan \Psi \cdot \cos \lambda = \tan \Psi \cdot |\sin PA| \qquad (3)$$
$$\cos \Psi = \cos \Phi \cdot \cos \Lambda \qquad (4)$$

HS's procedure uses the latitude and longitude as independent parameters. However, they cannot be freely selected for a given event because each ICME has a position angle determined from the SMEI images. It reduces selection to the pairs ($\Lambda,\Phi$) that satisfy Equation (1). The Cube-Fit procedure does not do this. Let's take some examples from HS's Tab. 1.

event # 004:

The procedure determined the direction to be S75 E75 (third column of the table). From Equation (1) it follows that $\lambda = 75.5°$, and thus for the SE quadrant PA = 165.5°; whereas the SMEI PA was 106° (second column in the table), that is ~60° lower.

event # 021:

The direction was determined as N00 E90 for this event. It corresponds to PA = 90°. The SMEI PA was 20°, that is 70° lower.

event # 056:

The determined direction is N00 W60, thus PA is 270°; while the SMEI PA was 357°, i.e. 87° greater.

Moreover, the use of two parameters of the direction is needless. Actually, the elongation-time plots are dependent on merely one parameter of the direction, the center disk distance. It follows that the three-dimensional Cube-Fit procedure should be replaced by a Square-Fit procedure with two free parameters, the speed (V) and the center disk distance ($\Psi$) as a parameter of the direction. Then the latitude and longitude might be determined from the assigned $\Psi$ and the measured SMEI PA using Equations (2) and (3). The determined direction would be consistent with the measured position angle; besides, the procedure would be largely simplified.

Another disadvantage of using the latitude and longitude as parameters of the model is that values of the center disk distance are unevenly and wastefully represented in the data cube.

Tab. 1 provides the center disk distances for the latitudes and longitudes employed in the data cube. They are determined using Equation (4). As one can see, there are many doubled values, and by implication, the data-cube plots corresponding to them are identical. For example, a value of 90º repeats 14 times. In total, about half of the plots is doubled, so they are needlessly checked by the procedure.

Furthermore, the values of Ψ are unevenly distributed; distances between them range from less than 0.1º up to 10º, and they became more dense as Ψ increases. This introduces bias in determined speeds, as it is demonstrated in the next section.

### 3. Broad set of plots instead of one plot matching a given event

A shape of an elongation-time plot depends on two free parameters: the speed and direction of ICME. This involves that there might be more then one plot matching a given event with the same accuracy. To demonstrate it let's consider for example a fictitious, but typical, event where elongation rises from 40º to 60º during 10 hours.

Fig. 2 shows a bunch of various plots matching the data with an accuracy of 0.5º. The accuracy is defined as the upper limit to the average model-data differences. Parameters of the plots (speed and direction), and time onsets (with respect to the first SMEI observation) range from 200 km/s, 5.5º, and -188 hours (top outermost plot) to 2800 km/s, 84º, and -11 hours (bottom outermost plot), respectively.

Fig. 3 shows the map of (V,Ψ) pairs for which the modeled plots match the event with an accuracy of 0.25º, 0.5º, 1º, and 2º (the innermost, two middle, and the outermost contour line, respectively).

The contours in Fig. 3 are situated diagonally, with increasing V as Ψ increases. This tendency is understandable if you take into account that a slope of the elongation-time plots gets more steep when speed is greater and center disk distance is smaller (the latter is true for elongations at which ICMEs are observed, i.e. from ~20º to ~140º ). It follows that in order to keep the data-point slope, $\Psi$ should be larger for larger V.

As one can see in Fig. 3, if higher accuracy is imposed the contours get narrower and they become a little shorter. However, there is still a broad range of speeds and center disk distances that match the data. This range cannot be freely diminished because a lower limit to the accuracy is imposed by the accuracy of elongation determinations made with SMEI.

Thus there is always many plots well matching a given event that should be considered when LASCO counterparts are searched for. The Cube-Fit procedure selects only one of them, not necessarily the best, because merely the plots included in the data cube are checked. A grid of points corresponding to the data-cube plots is shown in Fig. 3. As it was noted in the previous section, the points become closer for higher $\Psi$. As a result, plots for large $\Psi$ (and thus implicitly large V) have greater chance to hit the best-fit area, so, to be chosen.

This partly might explain why the distribution of the SMEI speeds determined by HS is excessively shifted toward high speeds (see HS, section 3, [18], and Fig. 2). Another cause is discussed in section 4, item 1.

Considerations of this section lead to the conclusion that an onset time of ICME cannot be unambiguously extrapolated from the SMEI data. Although the technique proposed by HS enables reduction of the range of admissible speeds and directions (differently in each case), this range is still broad. Consequently, also the range of the onset times is broad, much broader than the time-windows used by HS to search for LASCO counterparts. Therefore, it is necessary to reexamine the associations presented in HS's Tab. 1. This is of particular importance as concerned the 'weak associated' and 'missing LASCO data' events because the qualification might be changed for them. If you examined the LASCO data for long enough time-periods preceding the events qualified by HS as 'missing

LASCO data' you would probably find LASCO counterparts to such events. It follows than there is not any reason to assert now that some SMEI ICMEs are undetected by solar coronagraphs.

**4. Other questions**

1. The Cube-Fit speeds given in HS's Tab. 1 (third column) are about twice as high as those demanded to obtain the onset times given in the second column of the table.

This can be easily demonstrated if you examine third event in HS's Fig. 3. The event has a number of #069 in HS's Tab. 1. The Cube-Fit speed and direction, listed in Tab. 1, are 1000 km/s and N30W60, respectively. The elongation-time plot for these parameters can be found in HS's Fig. 1b (third plot from the bottom). From the plot, it can be read off that an elongation of ~47.5º (corresponding to the first observation of the event) is achieved during ~ 32 hours, and an elongation of ~61º (corresponding to the last observation) is achieved after next ~ 12 hours. Respective time-periods in HS's Fig. 3 are ~65 and ~24 hours, that is nearly twice as longer.

This indicates that the Cube-Fit curve shown in HS's Fig. 3, as well as the onset time given in Tab. 1, correspond to a speed which is about a half of that given in the table. I examined several other events. In each case the onset time corresponded to a speed about twice as low as that listed in the table.

There is another evidence in favor of the above. All speeds in the table are even multiples of 100 km/s, while the speeds employed in the data cube are both even and odd multiples (see HS [12]). It is improbable that the procedure has selected an even multiple in each case. This indicates that the speeds in the table have been doubled by mistake. Since the speeds given in Tab. 1 are overstated, no wonder that their distribution, shown in HS's

Fig. 2b, is excessively shifted toward high speeds (HS, [18]).

2. HS list a zero speed as one of the parameters employed in the data cube (see HS, [12]). When the speed is 0 elongation is ~0º independently of time, so the model-data differences would be always greater than 17º (the lowest elongation for SMEI), even if the event had very low speed. Thus the zero-speed plots have not any chance to be selected, so, they are useless.

3. HS assigned the best-fit plot in two ways: selecting the plot for which sum of the model-data differences is smallest, and the plot for which average of the model-data differences is smallest (see HS, [12]). However, if the sum is smallest, the average is also smallest; so there is no need to use both ways.

Table 1. Center disk distances for the latitudes and longitudes employed in the HS's data cube.

|           | Latitude |      |      |      |      |      |      |
|-----------|------|------|------|------|------|------|------|
| Longitude | 0    | 15   | 30   | 45   | 60   | 75   | 90   |
| 1         | 1.0  | 15.0 | 30.0 | 45.0 | 60.0 | 75.0 | 90.0 |
| 5         | 5.0  | 15.8 | 30.4 | 45.2 | 60.1 | 75.1 | 90.0 |
| 15        | 15.0 | 21.1 | 33.2 | 46.9 | 61.1 | 75.5 | 90.0 |
| 30        | 30.0 | 33.2 | 41.4 | 52.2 | 64.3 | 77.0 | 90.0 |
| 45        | 45.0 | 46.9 | 52.2 | 60.0 | 69.3 | 79.5 | 90.0 |
| 60        | 60.0 | 61.1 | 64.3 | 69.3 | 75.5 | 82.6 | 90.0 |
| 75        | 75.0 | 75.5 | 77.0 | 79.5 | 82.6 | 86.2 | 90.0 |
| 90        | 90.0 | 90.0 | 90.0 | 90.0 | 90.0 | 90.0 | 90.0 |

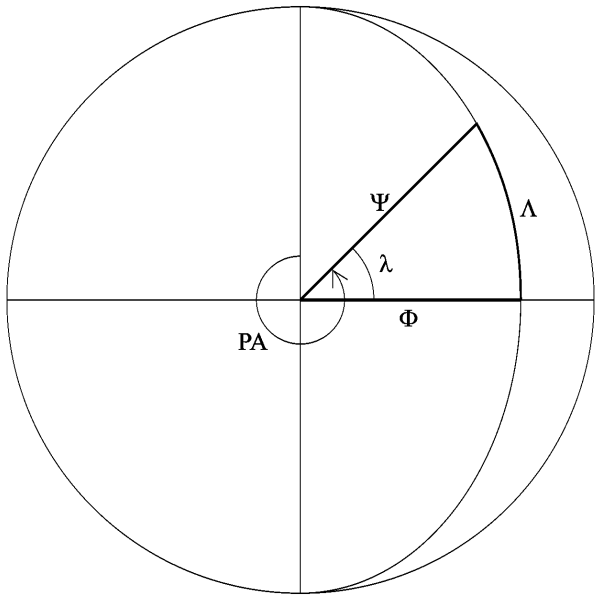

Figure 1. Geometrical relations between the parameters of the direction: latitude ($\Lambda$), longitude ($\Phi$), center disk distance ($\Psi$), position angle (PA), and apparent latitude ($\lambda$).

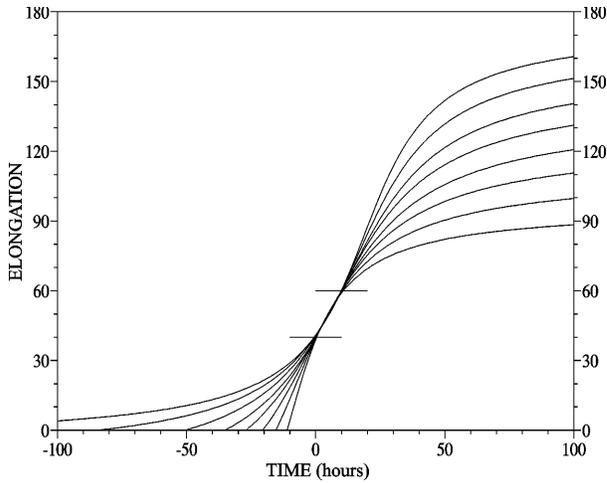

Figure 2. Bunch of the modeled elongation-time plots matching the fictitious event within an accuracy of 0.5°. Two horizontal lines mark the range of elongations corresponding to the event.

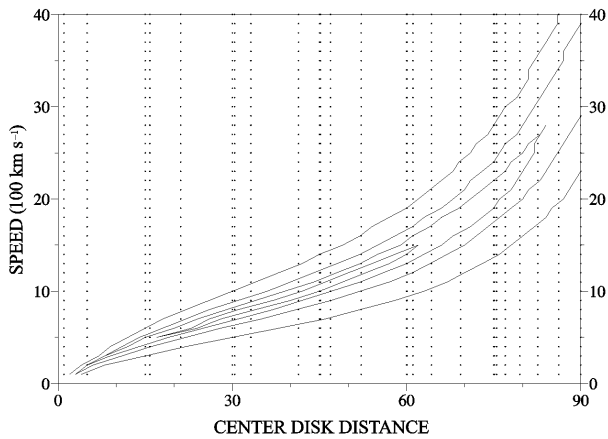

Figure 3. Map of the speeds and center disk distances for which the elongation-time plots match the fictitious event within an accuracy of 0.25°, 0.5°, 1°, and 2° (from the innermost to the outermost contour line, respectively). The point grid marks the parameters of the plots included in the HS's data cube.